\documentclass[[booktabs,reprint,groupedaddress,amsmath,amssymb,aps,prl,floatfix,longbibliography]{revtex4-1}

\usepackage{amsmath,amssymb,graphicx,mathrsfs,amsfonts,braket,dcolumn,bm,hyperref,color,braket,siunitx,setspace,xcolor}
\pdfoutput=1
\begin{document}

\title{A Robust Narrow-Line Magneto-Optical Trap using Adiabatic Transfer}

\author{Juan A. Muniz}
\affiliation{JILA, NIST, and University of Colorado, 440 UCB, Boulder, CO  80309, USA}
\author{Matthew A. Norcia}
\affiliation{JILA, NIST, and University of Colorado, 440 UCB, Boulder, CO  80309, USA}
\author{Julia R. K. Cline}
\affiliation{JILA, NIST, and University of Colorado, 440 UCB, Boulder, CO  80309, USA}
\author{James K. Thompson}
\affiliation{JILA, NIST, and University of Colorado, 440 UCB, 
Boulder, CO  80309, USA}
\email[]{juan.munizsilva@colorado.edu}


\date{\today}

\begin{abstract}
We characterize the properties of a new form of magneto-optical trap (MOT) that relies on non-equilibrium population dynamics associated with narrow-linewidth optical transitions. We demonstrate this mechanism using the 7.5~kHz linewidth transition in both bosonic and fermionic strontium isotopes.  In contrast to standard narrow-line MOTs based on imbalances in equilibrium scattering rates, our system benefits from a less complex laser system, a larger capture fraction from higher temperature samples, robustness against experimental perturbations and the ability to operate in the presence of large inhomogeneous broadening of the atomic transition frequency.
\end{abstract}

\maketitle


Laser cooling and trapping has been widely investigated since the advent of Doppler cooling in the 1980's \cite{chu1986experimental,Hansch_1975,Lett_89}. The first experiments used alkali atoms, with typical transition linewidths of a few MHz. A natural extension to achieve lower temperatures and high phase-space density is to use narrow-linewidth optical transitions, found for instance in alkali earth and rare-earth elements. Narrow-line transitions have proved interesting not only for laser cooling of atoms and molecules \cite{Katori_1999,Mukaiyama_2003,Loftus_2004_1,Loftus_2004_2,Bennets_2017,Collopy_2015} and quantum gas applications \cite{Stellmer_2009,deEscobar_2009,Duarte_2011,Kraft_2009,Stellmer_2013}, but also for precision measurements \cite{Meiser_2009,Meiser_2008,Salvi_2018,delAguila_2018} and optical atomic clocks \cite{Ludlow_2015}.

In this letter, we present a new procedure to generate a MOT using a narrow linewidth transition. This technique is motivated by the recent demonstration of Saw-tooth Wave Adiabatic Passage (SWAP) cooling of $^{88}$Sr atoms using the narrow-line 7.5~kHz optical transition \cite{NorciaSwap_2018,Bartolotta_2018}, and using Raman transitions to cool $^{87}$Rb atoms to temperatures below the Doppler cooling limit \cite{Greve_2018}.

The SWAP MOT configuration is the same as that of a standard MOT: three sets of counter-propagating laser beams with opposite circular polarization intersect at the center of a magnetic quadrupole field \cite{metcalf2007laser}. Unlike a standard MOT, our technique relies on adiabatic transfer between the ground state and one of two long-lived optically excited states to both cool and confine the atoms. The SWAP MOT is created by sweeping the frequency of the MOT beams upwards across the optical transition in an asymmetric sawtooth manner. The presence of Zeeman shifts on the long lived excited states create an imbalance in the number of photons absorbed from each beam, resulting in a magnetic field dependent restoring force. Compared to more traditional narrow-line MOTs \cite{Mukaiyama_2003}, our system benefits from a simpler laser system, larger capture fraction from higher temperature samples, improved robustness, and the ability to operate in the presence of large inhomogeneous broadening of the atomic transition frequency.

In contrast to SWAP cooling, where only occasional spontaneous emission is required, our SWAP MOT relies heavily on spontaneous emission to reset the atom to its ground state at the beginning of each sweep. However, the magnitude and spatial range of the force are both larger than those of the radiation pressure forces used for a standard MOT. Other works have previously observed enhanced optical forces by using adiabatic transfer techniques to cool and deflect atomic samples \cite{Voitsekhovich_1988,Nolle_1996,Sloding_1997,Miao_2007,Stack_2011}, but our work presents a simpler mechanism to provide both trapping and cooling using adiabatic transfer.


\begin{figure}[!htb]
\includegraphics[width=3.375in]{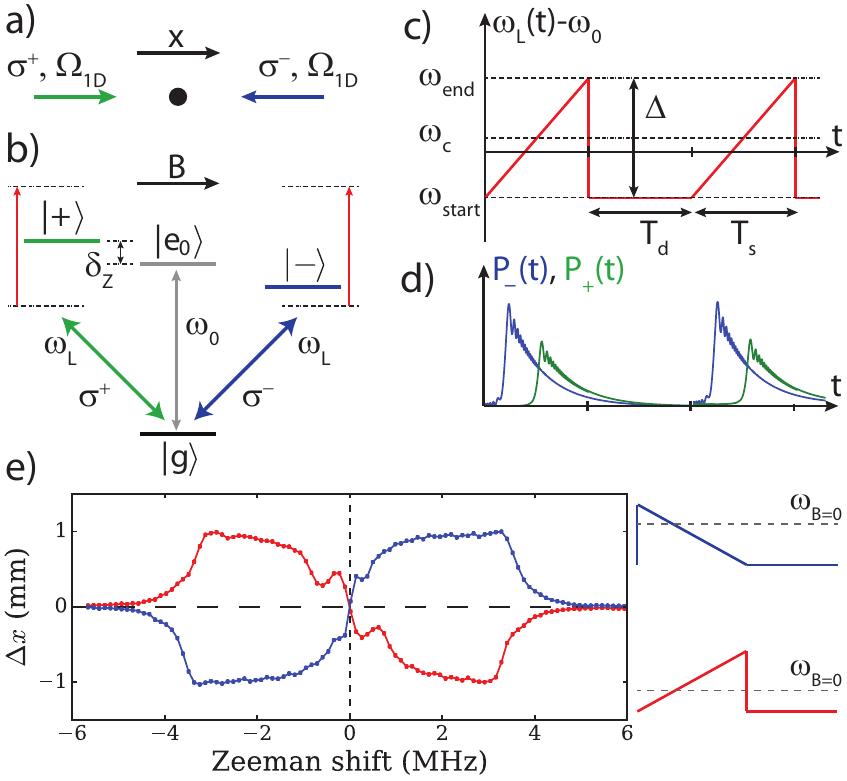}
\caption{SWAP MOT restoring force mechanism in 1D. (a) A pair of counter-propagating and oppositely polarized laser beams interact with an atom in a region with magnetic field $\vec{B} = B\hat{x}$. Each laser has Rabi frequency $\Omega_{1D}$. (b) The atom has a single ground state $\ket{g}$ and excited states $\lbrace\ket{-},\ket{e_{0}},\ket{+}\rbrace$. In the presence of a magnetic field, each excited state is shifted by $\delta_{\textrm{Z}}\propto m_JB$. (c) Laser frequency sweep $\omega_L(t)-\omega_{0}$. The start and end sweep frequencies relative to $\omega_0$ are $\omega_\textrm{start}$ and $\omega_\textrm{end}$, respectively. (d) Probabilities $P_{+}(t)$ and $P_{-}(t)$ of being in the respective excited state, as $\omega_L(t)$ sweeps across both transitions, as in (b). (e) Displacement $\Delta x$ after time of flight for downward (blue) and upward frequency sweeps (red). Both have $(\Delta/(2\pi),T_s,T_d,\Omega_{1D}/(2\pi)) = (4$~MHz$,~66~\mu$s$,~66~\mu$s,$~470$~kHz), while $\omega_c=2\pi\times1.45~(-1.45)$~MHz for the blue (red) points}
\label{fig:introd}
\end{figure}

To study the mechanism behind the restoring force, we first perform a simpler experiment in 1D, in the presence of a uniform magnetic field $\vec{B} = B\hat{x}$ and a pair of counter-propagating lasers along the $x$ direction (quantization axis), with equal Rabi frequency $\Omega_{1D}$ as in Fig.~\ref{fig:introd}(a). We consider a simple atomic structure with a ground state, $\ket{g}$, connected to a trio of optically excited states, $\lbrace \ket{-},\ket{e_0},\ket{+}\rbrace$, via a narrow-line transition, as in Fig.~\ref{fig:introd}(b). The energy difference between $\ket{g}$ and $\ket{e_0}$ at zero magnetic field is $\hbar\omega_{0}$. Both laser beams have the same frequency, $\omega_L(t)$, swept in an asymmetric saw-tooth fashion with period $T_s$, center frequency $\omega_{c}$, relative to $\omega_{0}$, and sweep span $\Delta$, as depicted in Fig.~\ref{fig:introd}(c). We typically operate with $T_s\geq 1/\Gamma$. In order to ensure that at the beginning of each sweep the atoms start most likely in the ground state, we allow a dead time $T_d$ between each sweep ($T_d\gg 1/\Gamma$), where the lasers are turned off. 

In this experiment, we use $^{88}$Sr, with ground state $^{1}\textrm{S}_0$ and excited states $\lbrace \ket{-},\ket{e_0},\ket{+}\rbrace$ in the $^{3}\textrm{P}_1$ manifold, labelled according to their angular momentum projections $m_J= \lbrace-1,0,+1\rbrace$, respectively. The excited state linewidth is $\Gamma = 2\pi\times$ 7.5~kHz and the transition wavelength is $\lambda = 689$~nm, with associated wavevector $k=2 \pi/\lambda$. In the presence of the magnetic field $B$, the excited states $\ket{\pm}$ experience a Zeeman frequency shift $\delta_{\textrm{Z}}/(2\pi) = m_J B\times 2.1~$(MHz/G).

In order to measure the momentum imparted to the atoms, we sweep the lasers fifteen times, then measure the displacement $\Delta x$ of a cloud of pre-cooled atoms after a brief time of flight. Figure \ref{fig:introd}(e) shows the displacement versus the Zeeman shift induced by the applied magnetic field for two different ramp directions. Noticeably, the sign of the displacement, and hence the effective force, flips depending on the frequency ramp's direction.
  
The force is generated when the laser is swept over the Zeeman shifted states. To describe the basic mechanism, we will assume that the sweep is configured as in Fig.~\ref{fig:introd}(c), that the Landau-Zener condition is satisfied \cite{Landau1932,Zener1932}, that at the beginning of each sweep the atom starts in $\ket{g}$, and that the Zeeman shifted transitions are in between $\omega_\textrm{start}$ and $\omega_\textrm{end}$, as in Fig.~\ref{fig:introd}(b). When the laser sweeps over the transition between $\ket{g}$ and the excited state that is shifted to lower energy, it adiabatically transfers the atom to this state and exchanges one photon recoil momentum, $p_r = \hbar k$, along the beam propagation direction. If the time before the laser frequency sweeps over the transition between the ground state and the higher energy excited state is short compared to the excited state lifetime $(2\delta_\textrm{Z}/\alpha\lesssim 1/\Gamma)$, it is likely that the atom will remain in the lower energy excited state when the laser sweeps across the higher frequency transition, preventing transfer to the higher-energy excited state. This creates an imbalance in the momentum exchanged between the atom and the two laser beams, as represented in Fig.~\ref{fig:introd}(d). 

To quantify this imbalance, we define $N_r$ as the total number of photon recoils acquired by the atom along $\hat{x}$ in a single sweep, such that after each sweep the atom's momentum changes by $N_rp_r$. This leads to a force $F = N_r p_r/T_s$ that can be measured through the cloud displacement. Because $N_r$ depends on the local magnetic field, in the presence of a quadrupole magnetic field, this force can be used to spatially confine the atoms at the minimum of the magnetic field.


\begin{figure}[!htb]
\includegraphics[width=3.375in]{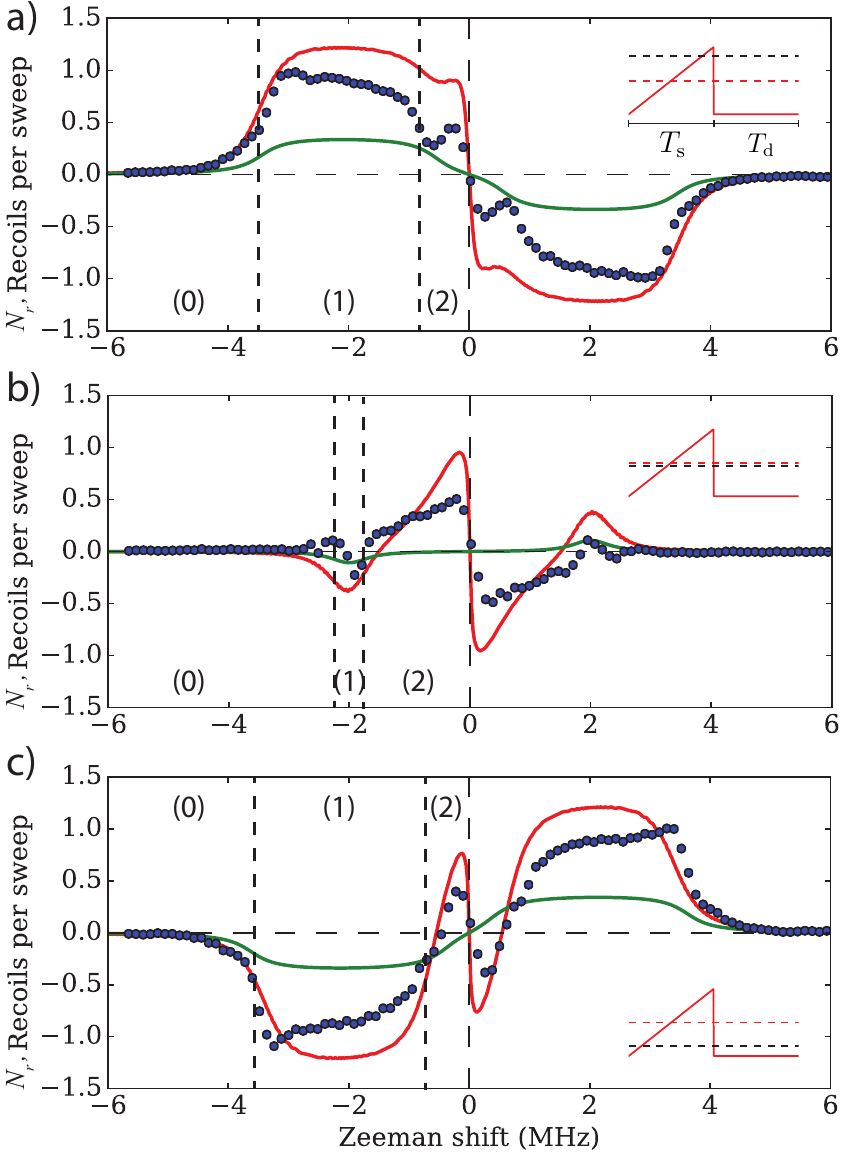}
\caption{1D restoring force curves. Blue points are the measured number of photon recoil momentum exchanged per sweep, $N_r$, for different applied Zeeman shifts ($\delta_\textrm{Z}$), red traces are the expected results based on numerically integrating the OBE for one sweep, while the green traces are based on the steady state scattering rate from each excited state. Each measurement was carried out with: $\Delta = 2\pi\times4$~MHz, $T_s=T_d=66~\mu$s, $\Omega_{1D} = 2\pi\times 470$~kHz, while the beams were applied for $T_p = 2$~ms and left to freely flight for $T_F = 10$~ms. In each plot (a,b,c), an inset shows $\omega_L(t)$ with respect to $\omega_{0}$ (black dashed line), for different center frequencies (red dashed line) $\omega_c = 2\pi\times(-1.45,0.15,1.62)$~MHz. Regions (0), (1) and (2) as in the text are indicated in each panel. Error bars $<0.02$.}
\label{fig:force curves}
\end{figure}

In Fig.~\ref{fig:force curves}, we measure $N_r$, as $B$ changes magnitude and sign, for different sweep center frequency $\omega_c$. To understand the features present in each force curve, we identify three different regions in each panel: (0), (1) and (2), named for the number of excited states crossed in a sweep. The first region with $N_r\sim0$, (0), occurs for values of $\delta_{\textrm{Z}}$ that lie outside the laser frequency sweep range $[\omega_\textrm{start},\omega_\textrm{end}]$. The second region, (1), is found for values of $\delta_\textrm{Z}$ where the laser only sweeps over one excited state, exchanging momentum only with the laser beam that addresses that state. In (a) and (c), region (1) manifests as a  non-zero plateau, because $\omega_c$ is large, whereas in (b), region (1) is a narrow region because $\omega_c$ is almost 0. Finally, the third region, (2), where $N_r$ rolls down to 0 at $\delta_{\textrm{Z}}=0$, arises when the sweep crosses over the two excited states. In that case, the atom can potentially exchange momentum with both beams, reducing the magnitude of the force. The transition width between each region is determined by $\Omega_{1D}$.   

To model the expected value of $N_r$, we numerically integrate the optical Bloch equations (OBE) for this three level system, $\left\{\ket{g},\ket{+},\ket{-}\right\}$, driven by two semi-classical fields with Rabi frequency $\Omega_{1D}$ and population decay from the excited states at a rate $\Gamma$ to obtain the density matrix $\hat{\rho}$. For each of the curves in Fig.~\ref{fig:force curves}, we use the experimental parameters to compute the excited state probabilities, $P_{+}(t)$ and $P_{-}(t)$, during a single sweep. As the scattering rate from each state is $\Gamma P_{\pm1}(t)$, the total number of recoils is $N_r = -\Gamma \int_0^\infty (P_{-}(t)-P_{+}(t))dt$, according to the sign convention adopted here. The results are shown as red curves in each panel in Fig.~\ref{fig:force curves} with qualitative agreement in every region. We attribute the quantitative disagreement between the measured and predicted value of $N_r$ to several experimental effects, such as imperfect polarization and power balance. Another important effect that seems to explain part of the roll off on the observed force as $\delta_{\textrm{Z}}$ approaches 0, is the presence of phase noise on the laser. 

The mechanism behind the SWAP MOT relies on the fact that the atomic populations do not equilibrate immediately to values that one would expect from the instantaneous laser detuning and intensity. To illustrate this, we calculate the force resulting from the steady state ($\dot{\hat{\rho}}(t)=0$) scattering rates from each excited state as the frequency sweeps, as in a standard MOT (Fig.~\ref{fig:force curves}, green curves) \cite{metcalf2007laser}. Clearly, the steady state behaviour does not capture many of the features present in the data, especially in regions where the sweep crosses both Zeeman shifted states (region (2)). 

For $\delta_\textrm{Z}=0$, an atom in the ground state is equally likely to interact with any of the beams, causing no net force ($N_r=0$). The tie in momentum exchange can be broken once the frequency splitting of the two excited states is nonzero. The size of this splitting depends on the laser field amplitude as well as its frequency, as it dresses the atom. We find that the force starts to roll off when the dressed state frequency shift, $\delta_\textrm{dr} \approx (-\omega_\textrm{end}+\sqrt{\omega_\textrm{end}^2+2\Omega_{1D}^2})/2$, is larger than $\delta_\textrm{Z}$. 

To form a MOT, one requires not only confinement, but also cooling.  This naturally occurs in the current configuration when the atoms are close enough to the zero of the quadrupole magnetic field such that the Zeeman shifts
$\delta_\textrm{Z}$ are much smaller than the Doppler shifts of the laser beams, $\delta_\textrm{D} = k v$, for an atom with speed $v$ along $\hat{x}$. In the reference frame of the atom, the laser beam counter propagating to the atom's velocity will appear $kv$ higher in frequency and the co-propagating beam will appear $kv$ lower in frequency. These Doppler shifts determine the time-ordering of absorption from the two beams in the same way that Zeeman shifts do at larger field values, providing damping of the atom's velocity.  Although it is possible to create a restoring force with downwards frequency sweeps by reversing the sign of the magnetic field as seen in Fig.~\ref{fig:introd}(e), only an upward frequency sweep enables a stable MOT with both confinement and cooling.

\begin{figure}[!htb]
\includegraphics[width=3.375in]{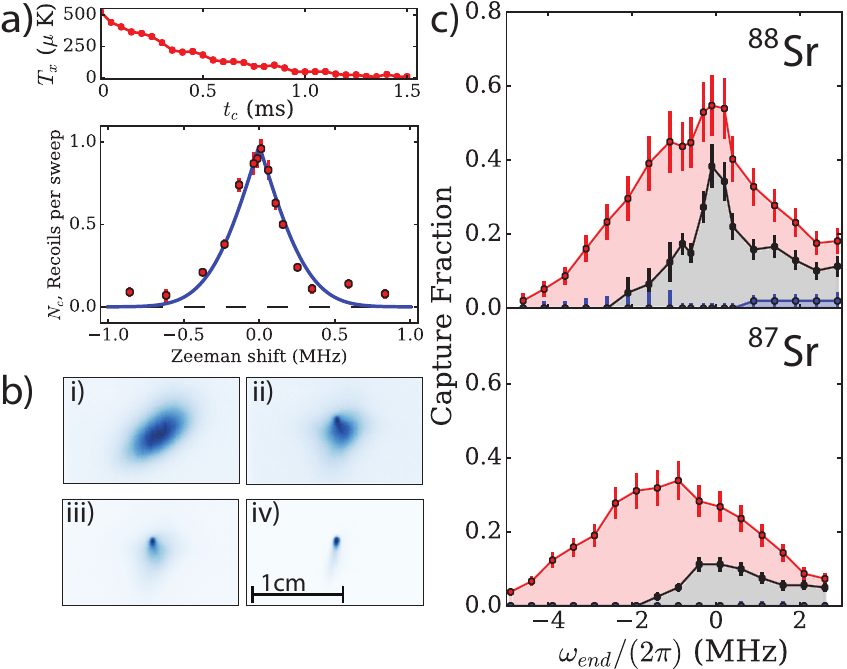}
\caption{Cooling and capture in the SWAP MOT. (a) Cooling in 1D. Top panel: $T_x(t_c)$ for $\delta_\textrm{Z}= 0$, for a cloud starting at 500$\mu$K. Lower panel: number of exchanged photon recoils per sweep, $N_c$ vs. $\delta_\textrm{Z}$ (red points), and 1D fit assuming cooling happens only for $\delta_\textrm{D} > \delta_\textrm{Z}$ (blue curve). (b) Fluorescence images at time $t_R$ during the blue to 3D SWAP MOT recapture process. For images (i,ii,ii,iv): $t_R = (0.5,20,30,150)$~ms, $N_\textrm{atoms} = (5.8,3.2,3.2,2.8)\times10^7$~atoms, $T = (750,135,43,37)\mu$K and $\rho_0 = (0.05,0.05,0.2,6.7)\times10^{10}$~at/cm$^3$. At the blue (SWAP) MOT the gradient is about 35~G/cm (5G/cm). c) Fraction of atoms transferred from the blue MOT to the SWAP MOT, for different $\omega_c$ (plotted against $\omega_\textrm{end}$). Red points represent an upwards saw-tooth ramp, black points represent a symmetric triangle frequency ramp, and blue points represent a downwards saw-tooth ramp. All the sweeps have $\Delta = 2\pi\times 3.3$~MHz, $T_s = 50~\mu$s, $T_d = 0$, $\Omega_{3D} = 2\pi\times300$~kHz.}
\label{fig:mot}
\end{figure}

To characterize the cooling mechanism and its modification due to a finite magnetic field, we apply the same set of beams for some time $t_c$ to a sample at around $500~\mu$K. We use time of flight experiments to measure the temperature of the cloud along $x$ as a function of the cooling time, $T_x(t_c)$, as shown in the top panel of Fig.~\ref{fig:mot}(a). If the atom's momentum changes by $-N_cp_r$ per sweep due to this force, the cooling rate at $t_c=0$ is $\dot{T}_x(0) = -N_c\sqrt{8T_\textrm{rec}T_x(0)}/(T_s+T_d)$, where $T_\textrm{rec} = \hbar^2k^2/2mk_B$ is the recoil temperature. We extract $N_c$ for different values of the applied magnetic field $B$, as displayed in Fig.~\ref{fig:mot}(a). The data suggests that cooling happens close to $\delta_\textrm{Z} = 0$, as anticipated before. Assuming cooling only happens when $\delta_\textrm{D} > \delta_\textrm{Z}$, a simple 1D Maxwell-Boltzmann distribution model suggests an initial sample temperature of $370\pm70~\mu$K, close to the measured value, as shown in the blue fit curve in Fig.~\ref{fig:mot}(a).

Now we will turn our attention to the 3D SWAP MOT. Even though this work refers mostly to measurements performed on the most abundant Sr bosonic isotope, $^{88}$Sr, similar results are found using the fermionic isotope $^{87}$Sr, with the laser addressing the $^{1}\textrm{S}_0$, $F=9/2$ to $^{3}\textrm{P}_1$, $F=11/2$ transition. Most experiments using standard radiation pressure forces to form a narrow-line MOT in $^{87}$Sr use a separate pair of beams, the so-called stirring beams, addressing the $F=9/2$ excited state manifold \cite{Mukaiyama_2003}, and/or frequency broaden their lasers matching the frequency spectrum to the velocity distribution of the atoms to improve the efficiency. Our scheme does not require a stirring beam as the sweep addresses the different shifts experienced by the atoms.

Figure \ref{fig:mot}(b) shows a few fluorescence images of the transfer from the blue MOT to the SWAP MOT, at the end of which, the phase space density is increased by a factor of $10^4$. We explore the efficiency with which we can recapture atoms for both isotopes $^{88}$Sr and $^{87}$Sr in Fig.~\ref{fig:mot}(c). Different frequency sweep profiles are applied: upward sweep (red points), downward sweep (blue points) and symmetric triangle sweep (black points), while the center frequency of the sweep, $\omega_c$, is changed. From now on, we set the dead time $T_d=0$. We observe that we capture substantially more atoms with the upward sweep, and over a broader range of frequencies. The improvements are more pronounced in our system for $^{87}$Sr than for $^{88}$Sr. Similar to the 1D case shown in Fig.~\ref{fig:mot}(a), we confirmed that a critical condition to have efficient cooling is that the laser sweeps over $\omega_{0}$, properly addressing the Doppler shifts near the quadrupole's center, as reflected in Fig.~\ref{fig:mot}(c).


\begin{figure}[!htb]
\includegraphics[width=3.375in]{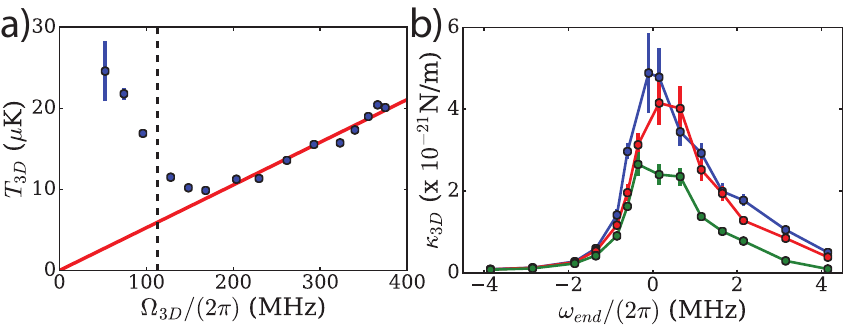}
\caption{3D SWAP MOT. (a) Steady state temperature as function of $\Omega_{3D}$. The black dashed line indicates the critical value, $\Omega_{cr}$, for adiabatic transfer to be successful. The red line is a linear fitting for $\Omega_{3D}>\Omega_{cr}$. (b) $\kappa_{3D}$ vs. $\omega_\textrm{end}$. Each sweep had $\omega_\textrm{start} = -2\pi\times4.85$~MHz, $T_s = 50~\mu$s, $T_d=0$. For the blue, red and green sets $\Omega_{3D} = 2\pi\times(380,320,240)$~kHz respectively.}
\label{fig:3Dmot}
\end{figure}

In Fig.~\ref{fig:3Dmot}(a), we study the MOT's steady state temperature as a function of the Rabi frequency, now taking into account the power in the six beams ($\Omega_{3D} = \sqrt{6}\Omega_\textrm{1 beam}$). Above the adiabatic transfer threshold, $\Omega^2_{cr} = \alpha$, the temperature rises proportionally with $\Omega_{3D}$. We find experimentally that $T_{3D} = n \hbar\Omega_{3D}/k_B$ with $n =1.10\pm0.02$, reflecting similarities with the mechanism behind SWAP cooling \cite{NorciaSwap_2018,Bartolotta_2018}. Below $\Omega_c$, the temperature starts to rise and the atom number in the SWAP MOT decreases quickly. Temperatures as low as  $10\mu$K are reached. 

For a wide range of experiments, we use the SWAP MOT to load and cool atoms into an intracavity 1D lattice at the magic wavelength of the Sr clock transition ($\lambda_\textrm{magic}=813$~nm) \cite{Norcia_2016,Norcia_2017spin,Norcia_2017frequency}. Our lattices are relatively deep, about 1000$E_\textrm{rec}$, causing large differential shifts for the cooling transition $^1S_0$ to $^{3}P_1$. We have observed that SWAP MOTs can still load both $^{88}$Sr and $^{87}$Sr atoms into these deep lattices from the SWAP MOT with close to $50\%$ efficiency and robustness and final temperatures of 5 to 10~$\mu$K. 

Although the 1D position dependent force curves shown in Fig.~\ref{fig:force curves} have several features that distinguish them from a typical MOT force curve, they are quite linear near $\delta_\textrm{Z}=0$. For our typical steady-state SWAP MOT the Zeeman shift associated with the rms cloud size, $r_{rms}$, is $\delta^{rms}_\textrm{Z}\leq2\pi\times0.2$~MHz, well within the linear regime. Therefore, we assume that the potential experienced by the atoms is nearly quadratic and try to measure the associated spring constant, $\kappa_{3D}$. Based on the equipartition
theorem, $k_B T_{3D} /2= \kappa_{3D}r^2_{rms}/2$, a measurement of the cloud temperature and spatial extent can determine $\kappa_{3D}$. 

Figure \ref{fig:3Dmot}(b) shows measurements of $\kappa_{3D}$ as a function of $\omega_\textrm{end}$ for three values of $\Omega_{3D}$. When $\omega_\textrm{end}<0$, the atoms experience a box like potential, flat near $\delta_\textrm{Z}=0$ and with characteristic width $2\omega_\textrm{end}$. The measured $r_{rms}$ for this set of points is independent of $\Omega_{3D}$ and decreases linearly with $\omega_\textrm{end}$. As $\omega_\textrm{end}$ approaches 0, the spring constant quickly rises. For values $\kappa_{3D}\approx 5\times10^{-21}$~N/m, the oscillation period is around 35~ms, while typical cooling times are below 1~ms, giving rise to a strongly overdamped MOT. For positive values of $\omega_\textrm{end}$, the spring constant slowly relaxes, showing a stronger dependence on $\Omega_{3D}$, because both the restoring force and the temperature depend on it. 

We have reported and characterized a robust and experimentally simple narrow-line MOT in Sr. The technique is very easy to implement, requiring a simpler laser system and proves to be robust against variation of experimental parameters. We have also developed a simple model to explain the behaviour of the restoring force. Future work may look towards using this technique to cool and load atoms into optical dipole traps, as well as to effectively cool and compress atomic samples for transport between two different spatial regions for continuous superradiance and matterwave lasers \cite{Meiser_2009,Bohnet_2012steady,Norcia_2017frequency,Bloch_1999,Hagley_1999,Robins_2008}.  

\section*{ACKNOWLEDGMENTS}\label{ack}
  
This work was supported by NSF PFC grant number PHY 1734006, DARPA QuASAR and Extreme Sensing, and NIST.  J.R.K.C. acknowledges financial support from NSF GRFP.

\bibliographystyle{apsrev4-1}
\bibliography{main}

\end{document}